\documentclass[conference]{IEEEtran}
\usepackage{tcolorbox}
\usepackage{etoolbox}
\AtBeginEnvironment{tcolorbox}{\small}
\usepackage{blindtext, graphicx}
\usepackage[compatibility=false]{caption}
\usepackage{subcaption}
\usepackage[super]{nth}
\usepackage{multirow}
\usepackage{mathtools}
\usepackage{url}
\usepackage{enumitem}

\usepackage{cite}

\ifCLASSINFOpdf
\else
\fi
\usepackage{array}
\usepackage{fixltx2e}

\usepackage{stfloats}
\usepackage{url}

\begin{document}
%
\title{Less is More: Exploiting Social Trust to Increase the Effectiveness of a Deception Attack}

\author{\IEEEauthorblockN{Shahryar Baki}
\IEEEauthorblockA{Computer Science\\
University of Houston\\
Houston, Texas \\
sh.baki@gmail.com}
\and
\IEEEauthorblockN{Rakesh M. Verma}
\IEEEauthorblockA{Computer Science\\
University of Houston\\
Houston, Texas \\
rmverma2@central.uh.edu}
\and
\IEEEauthorblockN{Arjun Mukherjee}
\IEEEauthorblockA{Computer Science\\
University of Houston\\
Houston, Texas \\
amukher6@central.uh.edu}
\and
\IEEEauthorblockN{Omprakash Gnawali}
\IEEEauthorblockA{Computer Science\\
University of Houston\\
Houston, Texas \\
odgnawal@central.uh.edu}
}


%



\maketitle

\begin{abstract}
Cyber attacks such as phishing, IRS scams, etc., still are successful in fooling Internet users.
Users are the last line of defense against these attacks since attackers seem to always find a way to bypass security systems. Understanding users' reason about the scams and frauds can help security providers to improve users security hygiene practices. In this work, we study the users' reasoning and the effectiveness of several variables within the context of the company representative fraud. Some of the variables that we study are: 1) the effect of using LinkedIn as a medium for delivering the phishing message instead of using email, 2) the effectiveness of natural language generation techniques in generating phishing emails, and 3) how some simple customizations, e.g., adding sender's contact info to the email, affect participants perception. The results obtained from the within-subject study show that participants are not prepared even for a well-known attack - company representative fraud. 
Findings include: approximately 65\% mean detection rate and insights into how the success rate changes with the facade and correspondent (sender/receiver) information. A significant finding is that a smaller set of well-chosen strategies is better than a large `mess' of strategies. We also find significant differences in how males and females approach the same company representative fraud. Insights from our work could help defenders in developing better strategies to evaluate their defenses and in devising better training strategies. 
\end{abstract}


%
\IEEEpeerreviewmaketitle

\section{Introduction}
So far, security has been an arms race in which defenders are typically behind the attackers. Researchers have tried to counter this issue by publishing new attacks or vulnerabilities in software and hardware, yet people continue to succumb to ``old'' attacks. How can defenders leap ahead of the attackers is the key question. For this purpose, we need to understand: (i) how people reason about scams and attacks, (ii) what makes an attack successful, and (iii) how can attackers increase their chances of success. Armed with this knowledge, defenders can trace the evolution of attacks and use it to bolster their defenses further and also devise more effective training regimes.

Given the scale of threat, every IT organization uses two prong strategy in defense: use of software tools to identify and filter phishing emails even before they reach the mailboxes and educate the users to identify phishing emails. Still, phishing continues to be effective as it increases in sophistication to evade tools and training. Understanding what makes phishing emails effective is critical to building anti-phishing tools.
Understanding how different aspects of a phishing email contributes to deception and action by the victim is challenging. The content, form, and context surrounding the phishing emails constitute a large number of variables. Over the past decade, emails have grown in complexity and routinely include tens of images, tables, formatted paragraphs in different fonts, sizes, and colors. They now resemble a modern webpage with the exception of executable scripts, which are disabled by most email clients. There could be tens to hundreds of variables controlling the visual elements of the emails that appear to be sent by sites such as LinkedIn or Facebook or banks. There are simply too many individual variables to study in a traditional laboratory or limited live attack setting. Many studies arbitrarily select a few variables (font size, logo tampering, link placement, color, etc.) and study how participants use these selected visual aspects of email to identify phishing emails.

Phishing emails oftentimes use visual scaffolding mimicking emails sent from well-known online services to increase the credibility of the email. An email that convinces a victim to click on a link to reset the password not only contains the password reset link but is fully wrapped in the style (logo, color, text, and other formatting details) used by legitimate emails sent from the target website. The user thus is led to believe that the email is legitimate and clicks on the link, go to the website to provide old password, for example, and fall victim to phishing.

Selecting a few variables for a deeper study to understand how and if they are used by users to identify phishing emails makes practical sense. This practical approach allowed the researchers to make progress in understanding contributing factors to successful phishing and building better anti-phishing tools and training. Fortunately, emails of the past generation were mostly text with sparse use of images, links, and formatting. In that context, the practical approach that studied a few variables was the right approach. Unfortunately, the emails have evolved significantly in the last 5-10 years to go from a few to tens to hundreds of objects, but the phishing studies have not been able to keep pace. Moreover, previous work has focused too much on phishing attacks and not enough on other kinds of email scams and attacks. 

To gain understanding of the parameters contribution to deception and action by the victim, we conduct a within-subjects study in which we take a simple, well-known attack, viz., company representative fraud, and parameterize it with signals such as fake logos of varying subtlety, surrounding context (facade), etc. We also consider automatic generation of these attacks using natural language generation techniques such as those employed in the Dada tool. Our goal here is to study whether such techniques are ``mature enough'' so that they can be deployed in credible attack generation. Being able to synthesize new attacks allows defenders to test their techniques on such new attacks rather than testing their methods on classical attacks. 

We design the experiment so that the participants have ample opportunity to exhibit their behavior and their reasoning. Our experiment is divided into two phases. In the first, we ask participants whether they believe the company representative email is genuine or fake and explain all the reasons for their judgment. In the second, we embed signals in the email and tell the participants that the email is fake, but ask them to explain all the ``flags'' indicating the email is fake, so that we can see how good are the participants at detecting them and also check whether participants are revealing all their reasoning in the first section. Participants took a personality test before the experiment and a debriefing discussion after.
Our findings include:


\begin{enumerate}
	\item We study how complex visual facades can be easily used to generate realistic and personalized attacks. Our work suggests that social scaffolding such as a LinkedIn context increases the effectiveness of the attack more than email clients such as Gmail (Section~\ref{subsec-gmail-linkedin}).
	\item Our study shows that adding sender's contact info and receiver's name to the email can significantly increase the success rate of the attack (Section~\ref{subsec-attack-type}).
    \item The more the strategies used by participants the lower the detection rate. We analyzed the data in different ways and this finding persists. Moreover,  there are significant differences in the strategies used by males versus those used by females (Section \ref{sec-results}). However, we find there is no statistically significant difference in the performance of the two groups, which suggests that there are multiple sets of strategies leading to similar performance. 
    \item 91\% of the participants did not notice tampering in a widely known logo. Specifically, only two participants detected our easiest fake logo and only one detected the next easiest one. No one detected level 3 and level 4 (hardest to detect) fake logos (Section \ref{sec-reasoning}). This insight can be used to better train Internet users and employees.
    \item We also study whether Natural Language Generation (NLG) technology can be used to semi-automatically generate effective attacks  (Section \ref{subsec-nlg-human}). Our results showed that participants do a better job in detecting fake representative offers generated by NLG. We believe that more complex grammar may be needed than we have used, since offers tend to be long and complex.
\end{enumerate} 

As pointed out in \cite{kelleyB16}, some empirical studies focus on binary decisions (i.e., whether the subject used or did not use the signal), while others consider indirect metrics such as the time taken \cite{schechter2007emperor,alsharnouby2015phishing}.\footnote{We also note the use of 5-point Likert scale in some studies, e.g., \cite{wangHCVR12}.} Other researchers \cite{kelleyB16} have revealed the limitations of these methods.  In our study, we try to go deeper into the reasoning process employed by the subjects. This necessitates the use of a lab environment since a user is unlikely to have the patience to share reasoning strategies for every decision, even if only from a subset of decisions, made in the real world. Hence, some studies have tried to use the mouse movements as a proxy \cite{kelleyB16} while others have equipped the subjects with intrusive sensors such as EEG devices \cite{neupane2015multi}, which have their own limitations. 


\section{Facade Attack Exploiting Trust}
\label{sec-facade}
The attack that we introduce in this work can be considered as a deception attack with different ways of exploiting social trust, which we call {\em scaffolding attack}. Although there has been significant work in the context of phishing and spear phishing  \cite{wangHCVR12}, to our knowledge, nobody has investigated such a range of parameters, or how participants behave and reason about this attack in different scenarios. 

\subsection{Representative Offer}
The fraudulent company representative offers (representative offer for short) are mainly looking for representatives in other countries who will collect  money from customers in those countries. We used an online dataset of company representative fraud emails to collect them.\footnote{\url{http://www.419scam.org/419representative.htm}}. Those offers have some typos and grammatical mistakes. We do not fix these issue to discover whether participants pay attention to such details. Below is a sample representative offer:

\begin{tcolorbox}[size=small]
	Attention, 
	
    \bigskip
    
	I am Jeannie U. Ashe manager in Supreme Access Industrial Limited. 
	I owned a company that sells home storage solution, home décor and outdoor décor products in Turkey. We are searching for representatives who can help us establish a medium of getting to our costumers in Europe as well as making payments through you to us.
	Please contact us for more information. You will have to send in your credentials, so we can know your competence. Subject to your satisfaction you will be given the opportunity to negotiate your mode of which we will pay for your services as our representative in Turkey and Europe. 
	If interested forward to us your phone number/fax and your full contact addresses. We can assure you that this proposal is 100\% legal.
   
    Please if you are interested forward to this e-mail:Jeannie\_Ashe@SAI.com 
	
	1) your full names 
	2) phone number/fax 
	3) contact adresses, 
	4) any form of identity/international passport. 
    5) age 
	
	in this formal application a message will be sent to you as regards appointment.
    
    \bigskip 
	Thanks in advance.     
    
    \bigskip 
    
	Jeannie U. Ashe  
    
	Manager     
    
	Supreme Access Industrial Limited
\end{tcolorbox}

\subsection{Natural Language Generators and the Dada Engine}
One of the aims of computational linguistics and natural language generation (NLG) is to facilitate the use of computers by allowing the machine and their users to communicate using natural language. There are two main areas of research in this field: understanding and generation. Usually, a generator works by having a large dataset of knowledge to pull from that is then manipulated by programmable grammatical rules in order to generate readable text. This process is usually broken down into two steps: text planning, which is concerned with deciding the content of the text, and realization, which is concerned with the lexigraphy, and syntactic organization of the text. Some of the variations of these approaches appear in~\cite{galanis2009open}.  Moore and Swartout~\cite{moore1991reactive} present a top-down approach to text planning, which is responsive to some communicative goal a speaker has in mind. Information is selected from the knowledge base and organized into coherent text during the planning process. Paris~\cite{paris2013natural} presents a similar approach with additional constraints put on the lexical organization of the text. However, because of the top-down nature of these approaches, the planning is restricted from incorporating additional elements that the user might find useful. Hovy~\cite{hovy1987generating} presents an approach that allows for top-down and bottom-up approach simultaneously. Because of this increased flexibility, constraints become even more important. McCoy and Cheng~\cite{mccoy1991focus} present constraints that can be used to control this process. For our purposes, we used the Dada engine \cite{bulhak1996dada}, which is capable of both top-down and bottom-up approaches for email generation which is appropriate for our use. 

\begin{figure}[h]
	\centering
	\includegraphics[width=3.5in]{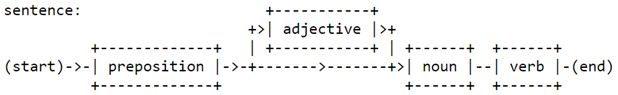}
	\caption{Example of Recursive Transition Network}
	\label{rtn}
\end{figure}

The Dada engine has been successfully used to construct the academic papers on postmodernism ~\cite{bulhak1996simulation}. The Dada engine works by pulling text from the knowledge base that is specified as appropriate by the constraints. The Dada engine is a natural language generator tool that is based on the principle of recursive transition networks or recursive grammars. A recursive transition network (RTN) can be thought of as a schematic diagram of a grammar which shows the various pathways that different yields of the grammar can take. For example, in the construction of a sentence, one may choose to follow the RTN shown in Figure ~\ref{rtn}.

If one follows the RTN in Figure~\ref{rtn} from the start to the end, one passes through states (boxes) representing the various elements which make up the sentence in sequence: first a preposition, then an optional adjective, then a noun and then a verb. After the preposition, the network branches into two paths, one of which leads to the adjective and the other which bypasses it and goes straight on to the noun. One may take either path in the course of constructing a sentence.

The building block of the NLG is a dataset of actual texts out of which we can extract the grammar rules. We used the dataset of company representative scams which is available online\footnote{http://www.419scam.org/419representative.htm} to create rules for Dada engine and also some of them were used directly in our representative offers.

\subsection{Hypotheses}
There are different clues in an email that can be a signal for users to decide whether an email is legitimate or fraudulent.  Being aware of those signals can help people to detect legitimate/fraudulent emails more easily. Hence, we pose the following hypotheses:

\begin{enumerate}[font={\bfseries},label=H1.\alph*., leftmargin=3.5\parindent]
\item Paying attention to the higher number of clues in the email can decrease participants vulnerability to representative offers.
\item People with higher knowledge about the email pay attention to more clues.
\end{enumerate}

Some social networks like LinkedIn are designed to be professional networks. Hence, people may trust message received from people in their professional network more than a email from a random person. This can be used by attackers to use professional network as a medium for delivering the message to the victims in order to improve the success rate of their attacks. Therefore, this study tests the effect of using professional social networks instead of traditional email-based phishing attacks by posing the following hypotheses:

\begin{enumerate}[font={\bfseries},label=H2.\alph*., leftmargin=3.5\parindent]
\item Using LinkedIn as a medium for delivering phishing emails improves the effectiveness of the attack compared to using email.
\item People with lower knowledge about the social networks are more susceptible to LinkedIn based phishing attack than people who are familiar with them.
\end{enumerate}

Representative offers are mostly used in mass distributions delivered to random recipient. For example, someone with a major in electrical engineering can receive a representative offer from a pharmaceutical company. This can be an obvious indicator for recipients that the email is fraudulent. In spear phishing attacks, attackers try to customize the message exactly for a specific person, but this necessitates a lot of knowledge about the victim and also lots of time to gather this information. In this study, we want to test how much some simple customization, like adding victims' name in the email, can increase the susceptibility of them. So, in this study we also pose the following hypotheses:

\begin{enumerate}[font={\bfseries},label=H3.\alph*., leftmargin=3.5\parindent]
\item Adding recipient's name in the greeting of the email decreases the detection rate of a fraudulent representative offer.
\item Adding the sender's signature to emails makes recipients feel the offer comes from a legitimate source. As a result, participants tag emails with signature as legitimate more frequently.
\end{enumerate}

Creating the representative offer emails is a time-consuming task. Using NLG tools can help to speed it up. This can be helpful for both attackers (generating a variety of emails in a shorter amount of time) and defenders (automatically generating datasets to improve existing detection systems). Hence, we present the following hypothesis to evaluate the effectiveness of the NLG tools:

\begin{enumerate}[font={\bfseries},label=H4., leftmargin=3\parindent]
\item The detection rate of fraudulent representative offers generated by NLG is not significantly different from human-generated offers.
\end{enumerate}

\subsection{Our study}
In this study, we investigate user behavior and reasoning with company representation emails. We parameterize the fraudulent emails with various ways of social trust exploitation. 

\begin{figure}[h]
	\centering
	\begin{subfigure}[b]{0.25\linewidth}
		\centering
		\includegraphics[width=.4\textwidth]{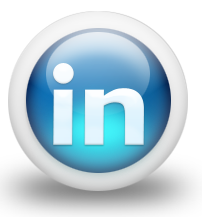}
		\caption{\centering{Level 1}}
	\end{subfigure}
	\hfill
	\begin{subfigure}[b]{0.25\linewidth}
		\centering
		\includegraphics[width=.4\textwidth]{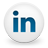}
		\caption{\centering{Level 2}}
	\end{subfigure}
	\hfill
	\begin{subfigure}[b]{0.25\linewidth}
		\centering
		\includegraphics[width=.4\textwidth]{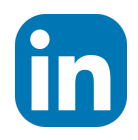}
		\caption{\centering{Level 3}}
	\end{subfigure}
	\hfill
	\begin{subfigure}[b]{0.2\linewidth}
		\centering
		\includegraphics[width=.4\textwidth]{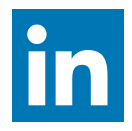}
		\caption{\centering{Level 4}}
        \label{fake-logo-level4}
	\end{subfigure}
	\caption{LinkedIn fake logos with various levels of subtlety}
	\label{fake-logo}
\end{figure}

\begin{enumerate}
    \item Using the context of LinkedIn or a traditional email-based attack (H2)
    \item With varying levels of contact information (sender only, recipient only, both sender and recipient, and none) (H3)
    \item With varying levels of customization using recipient's background information (education, work, or both of them)
	\item Obviously fake company names (e.g., \textit{Donald Duck and Mickey Mouse})
	\item Fake logos: There are four different fake logos with varing levels of subtlety (Figure \ref{fake-logo}).		
\end{enumerate}   
Some of the attacks are generated semi-automatically using NLG technology as employed in the DADA tool. We study whether these attacks are effective and how many participants are able to identify these ``synthetic'' attacks (H4).

\section{Experiment Setup}
\begin{figure}[h]
	\centering
	\includegraphics[width=3.2in]{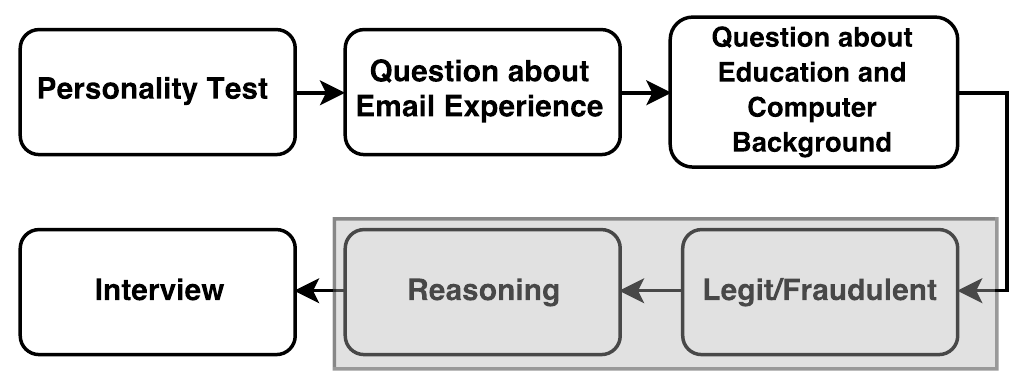}
	\caption{Flowchart of the entire experiment}
	\label{fig-flowchart}
\end{figure}

To evaluate the effectiveness of the attack under different scenarios, and to investigate the strategies that people use in their decision-making process, we perform a user study. Figure \ref{fig-flowchart} shows the diagram of the experiment. Before coming to the lab, all participants take a personality test. This is done to reduce the time spent in the lab  and the potential for fatigue. We used the Big Five Personality test \cite{john1999big} for measuring personality traits of participants. The Big Five Personality Traits is a well known and widely accepted model for measuring and describing different aspects of human personality and psyche~\cite{clark2007assessment}. This model describes human personality from five broad dimensions: Extraversion, Agreeableness, Conscientiousness, Neuroticism, and Openness. 
We created the personality test using Google Form\footnote{https://www.google.com/forms/about/} and asked the participants to do it before coming to the lab for the actual experiment. Table~\ref{table-personality-range} presents the range of possible values and range of the participants' scores in each trait. It shows that the participants' scores approximately covers from lowest to highest score for all the traits which means our samples from the population are not skewed in one or more directions from the personality point of view. 

\begin{table*}[h]
	\centering
	\caption{Range of possible scores in each trait of Big Five Personality test in addition to participants' scores}
	\label{table-personality-range}
	\begin{tabular}{|c|c|c|c|c|c|}
		\hline
		\multirow{2}{*}{Traits} & Minimum       & Maximum       & Mean    & Median  & StdDev  \\ \cline{2-6} 
		& Possible/Participants & Possible/Participants & Participants & Participants & Participants \\ \hline
		Extraversion            & 8/14          & 40/39         & 26.1    & 27.5    & 6.2     \\ \hline
		Agreeableness                  & 9/14          & 45/43         & 32.6    & 33      & 6.7     \\ \hline
		Conscientiousness               & 9/19          & 45/44         & 31.4    & 32      & 6.4     \\ \hline
		Neuroticism             & 8/11          & 40/38         & 24.2    & 23      & 6.1     \\ \hline
		Openness                & 10/21         & 50/46         & 35.2    & 35      & 5.6     \\ \hline
	\end{tabular}
\end{table*}

In the lab, we begin by asking participants a few basic questions about their email experience: approximate number of emails received each day, years of email use, and the spam filter used, if any. Next, we ask them about their education and computer background. These questions are followed by the scaffolding attack part of the experiment (gray part of Figure \ref{fig-flowchart}). 
It consists of two sections, the first one (Legit/Fraudulent) is designed to study the effectiveness of scaffolding, and the second (Reasoning) is intended to investigate in-depth the strategies employed by participants when the pressure to make a decision is relaxed.

In the Legit/Fraudulent section, we use two different contexts for sending emails to the recipients, Gmail and LinkedIn. There are two representative offers using Gmail and two using LinkedIn. One offer in each of them is generated by NLG (FakeN) and the other one is a `real' representative offer fraud (Fake). In the Email Delivered by Gmail (EDG), we show a screenshot of Gmail interface and ask participants to decide if this a legitimate email or fraudulent one. In the Email Delivered by LinkedIn (EDL), first, we show them a LinkedIn friend request (screenshot of a friend request email) and ask them to suppose after receiving the friend request they have received a message from that person via LinkedIn (screenshot of a LinkedIn message in their inbox). We also added six more representative offers consisting of both EDG and EDL. They are generated by using different combinations of information about the company that hires and the name of the recipient. To see how people perform in the case of emails without any fraudulent signals, we also use two real job offers in this section, but We change the LinkedIn logo in one of them with a fake logo (very similar to the actual LinkedIn logo, Figure \ref{fake-logo-level4}). 

For each attack scenario, we ask (i) ``Do you think this is a legitimate scenario or a fraudulent scenario?'' and (ii) ``If you think this scenario is fraudulent, please list ALL the reasons that made you think the email is fraudulent. Otherwise, justify why you think this scenario is legitimate.'' In addition to these questions, we also asked participants to indicate their confidence level, ``How confident are you about your answer?'' to know if they got lucky. Confidence ranges from 1 (least confident) to 5 (extremely confident). In order to remove the effect of the order of representative offers, we randomize them once at the beginning and then use the same generated order for all the participants. We do not change the order of questions for each participant since it affects their response to each question in a different way.

The Reasoning section consists of nine different representative offer, eight generated by the NLG and one offer with ``real'' representative offer fraud from the Internet. Out of those eight NLG generated offers, four have fake LinkedIn logos of different difficulty level (see Figure~\ref{fake-logo}), three are customized offers based on participants background, and one with an obvious fake company name. For the customized emails, customization is done based on: participants' field of study/work and the institution in which they studied. We gathered this information by asking the participants to send us their LinkedIn web page or their resume before coming to the lab. In this section, in contrast to the previous one, we do not ask participants to decide if the email/scenario is fraudulent or legitimate. We told them at the beginning that all the emails in this part are fake and they should just find the clues and factors that made this email/scenario fraudulent. 

At the end, we interviewed the participants to evaluate their knowledge and experience about the Email in more detail. ``What are the different parts of the email'' and ``How much do you rely on sender's email address in order to decide if an email is fake or not'' are some examples of these questions. We present the answers that we got from the interview in Section \ref{sec-interview}.


For running the survey, we utilized Form Maker\footnote{https://wordpress.org/plugins/form-maker/} which is a plugin for Wordpress,\footnote{https://wordpress.org/} a free and open source content management system. It enables us to keep track of the total time participants spent on reading each representative offer and the time they spent on answering questions. 

\subsection{Participants}
Before the experiment, we requested IRB approval. We also did a small pilot study with three participants before running the actual survey. During the pilot study, we found and fixed the following problems:
\begin{itemize}
	\item \textbf{Legitimate Emails}: We had used two emails from Enron Email Dataset. Since the topic of these emails did not match those of the Fake emails (representative offer), we replaced them with legitimate job offers.
	\item \textbf{Fake Company Name}:  The fake name that we used at beginning was hard to detect. So, we changed it to an obvious one (\textit{Donald Duck and Mickey Mouse}).
\end{itemize}

After IRB approval, a recruitment email was sent to all the students at the College of Natural Sciences and Mathematics, which includes six departments (and over 20 majors): Biology \& Biochemistry, Chemistry, Computer Science, Earth \& Atmospheric Science, Mathematics, and Physics. We also mentioned  in the email that participants will be given \$20 Amazon gift card upon finishing the experiment. To further diversify the participant pool, we also recruited staff, so we have some majors from other colleges also. We had 34 participants, of which 15 were female (44\%) and 19 male (56\%). The majors of our participants are computer science (33\%), Biology (26\%), Chemistry (12\%), finance (9\%) and others (20\%). From the academic degree aspect, four of them are Ph.D. students, two are Masters students, 26 Bachelors students and two of them have High School Diplomas.

Most of the participants use Gmail as their spam filter, 30 out of 34 (88\%). Kaspersky, Macafee, Yahoo, and Web Of Trust each is used by one participant. Below are some other statistics about the participants (since some of the data are skewed, we also report the median, as well as first and third quartile).

\begin{itemize}
	\item Age: Range $[18-64]$ years (\nth{1} quartile = 19, median = 21, mean = 25, \nth{3} quartile = 23, SD = 10.7).
	\item Number of emails received daily: Range $[3-100]$ (\nth{1} quartile = 7, median = 12, mean = 19.66, \nth{3} quartile = 20, SD = 19.74).
	\item Years of email usage: Range $[4-20]$ (\nth{1} quartile = 8, median = 10, mean = 10.55, \nth{3} quartile = 11.75, SD = 3.76).
	\item Social network usage (number of times checked per week): Range $[0-100]$ (\nth{1} quartile = 7, median = 20.5, mean = 71.81, \nth{3} quartile = 50, SD = 222.96).
\end{itemize}

\subsection{Variables}
\label{subsec-variables}
\begin{figure}[h]
	\centering
	\includegraphics[width=3.4in]{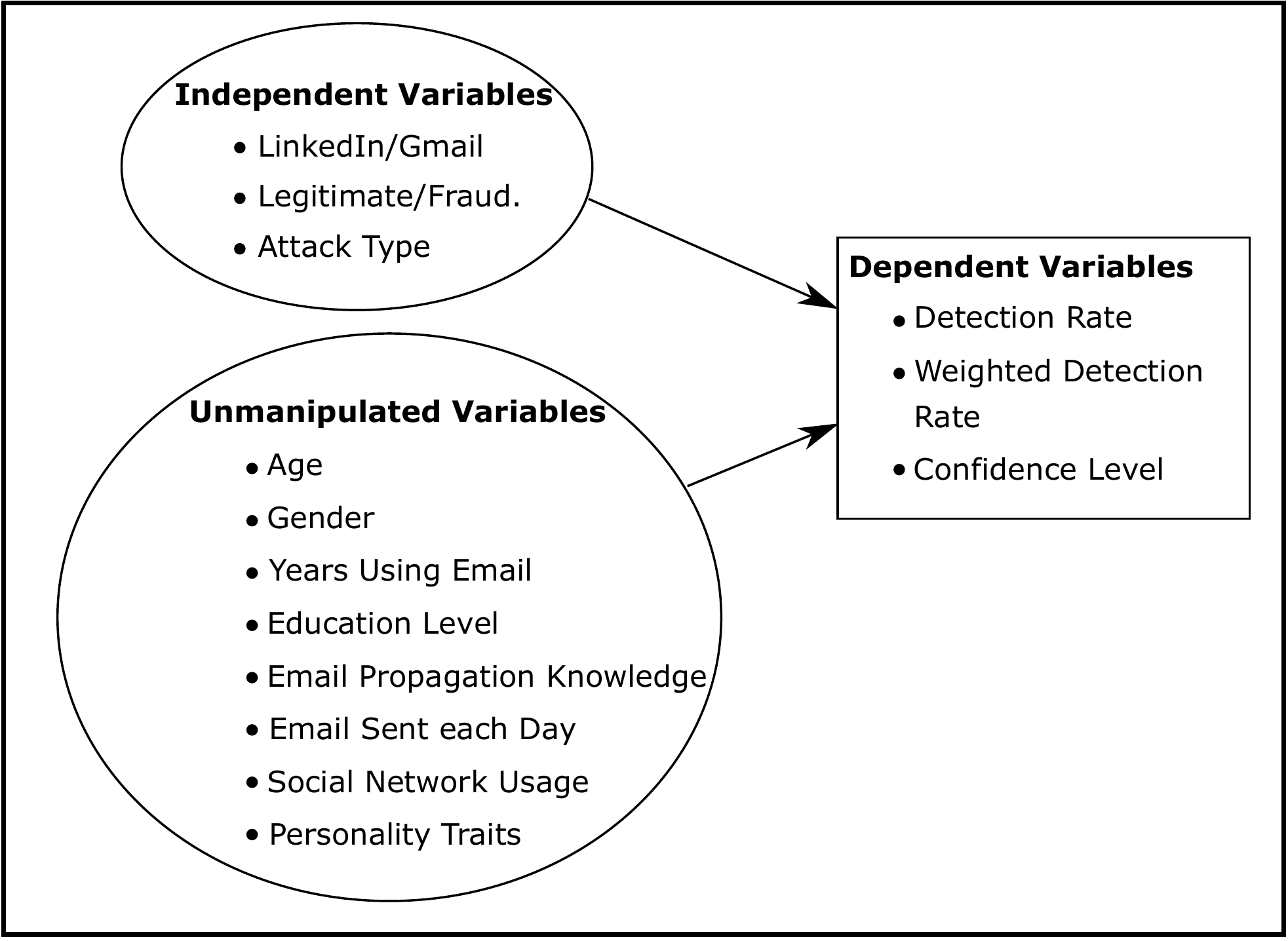}
	\caption{Relationship Between Variables}
	\label{fig-variables}
\end{figure}

Three categories of variables exist in the study: independent, dependent, and unmanipulated (extraneous) variables. Figure~\ref{fig-variables} shows the variables and their relationships. 
Independent variables are those that we manipulated on different representative offers to see how they affect participants' performance. Table~\ref{table-independents} lists and explains the independent variables.

Dependent variables are the observed variables and the goal of this study is to find how they depend on the other variables. Performance of participants on detection legitimate and fraudulent job offers is the main dependent variable in our study.
\textit{Detection rate} (DR) is the proportion of questions that are answered correctly, legitimate offer detected as legitimate and fraudulent detected as fraudulent. The \textit{confidence level} is the degree of confidence that participants have in their answers to the questions. In the result sections, we use the average confidence level (ACL) instead of separate confidence level for each question/participant. The third dependent variable is \textit{Weighted detection rate} which combines detection rate and confidence level together. DR only considers the final output of participants for each question, either zero or one. This combination (DR and confidence level) helps us to differentiate between participants who answered correctly with high confidence and low confidence, and conversely, those who answered incorrectly with high confidence and low confidence.


Equation \ref{equ-wdr} shows the relation of the WDR.

\begin{equation}
\label{equ-wdr}
\begin{aligned}
WDR &= \frac{\sum_{all\_questions} SDR*Confidence}{\left | all\_questions \right |}\; ,\; \\
SDR &= \left\{\begin{matrix}
1\; if\;  DR=1\\ 
-1\;  otherwise
\end{matrix}\right.
\end{aligned}
\end{equation}

\begin{table}[h]
	\centering
	\caption{Independent Variables and Description}
	\label{table-independents}
	\begin{tabular}{|c|c|}
    \hline
    Independent variable & Description                                                                                               \\ \hline
    LinkedIn/Gmail       & \begin{tabular}[c]{@{}c@{}}Medium used for sending representative\\ offer\end{tabular}                    \\ \hline
    NLG/Human            & \begin{tabular}[c]{@{}c@{}}Representative offer created by \\ NLG or human\end{tabular}                   \\ \hline
    Legitimate/Fraud     & \begin{tabular}[c]{@{}c@{}}Real job offer and fraudulent\\  representative offer\end{tabular}             \\ \hline
    Attack type          & \begin{tabular}[c]{@{}c@{}}Different attacks: Sender info., receiver info.,\\ both, and none\end{tabular} \\ \hline
    \end{tabular}
\end{table}

Unmanipulated variables are variables over which we have no control. By randomly choosing the participants from the population, knowing the fact that people reply the recruitment email randomly, we reduce the effect of these variables on our study. Nevertheless, we study if there is any relationship between them and dependent variables. Table~\ref{table-unmanipulated} lists and explains unmanipulated variables (age and gender are removed from the table). For \textit{Email Propagation Knowledge}, we define four levels: (i) zero knowledge, (ii) those who know there is a server that receives email from the sender and sends it to the receiver, (iii) those who know there is more than one server in the middle (sender email server and receiver email server), and (iv) those who have fairly comprehensive knowledge.

\begin{table}[h]
	\centering
	\caption{Unmanipulated Variables and Description}
	\label{table-unmanipulated}
	\begin{tabular}{|c|c|}
\hline
Unmanipulated variable      & Description                                                                                                  \\ \hline
Years using email           & Number of years using email                                                                                  \\ \hline
Education level             & Participants' seeking degree                                                                                 \\ \hline
Email propagation knowledge & \begin{tabular}[c]{@{}c@{}}Level of knowledge about how \\ emails are sent through the Internet\end{tabular} \\ \hline
Email sent each day         & Number of email sent each day                                                                                \\ \hline
Social network              & \begin{tabular}[c]{@{}c@{}}How many times per week \\ they use any social network\end{tabular}               \\ \hline
Personality traits          & \begin{tabular}[c]{@{}c@{}}Five personality traits defined\\ by Big Five Personality Test\end{tabular}       \\ \hline
\end{tabular}
\end{table}

\section{Performance on the Legitimate/Fraudulent Section}
\label{sec-results}
A common question that needs to be addressed in almost all user studies is ``do the predictor variables have any effect on dependent variable?'' In this section, we present the effectiveness of the attack scenarios that we introduce and whether any variables have any statistically significant effect on the performance of participants. We take a similar approach that has been used in \cite{alsharnouby2015phishing} to compare the results. We used R \cite{RCore} to perform all the significance and correlation tests.

First, we study the effect of independent variables which are the main goal of this experiment, and then the effect of unmanipulated variables. Table~\ref{table-leg/fraud-ques} shows details of each representative offer in the first section of the experiment with the performance of participants and their average confidence level. To cluster together offers that share a common variable, we changed their order in Table \ref{table-leg/fraud-ques} (the actual order is in Column 2). 

\begin{table}[h]
\centering
\caption{10 representative offers in the legitimate/fraudulent section (detection rate and average confidence level are included). Column. 1 -  variable changes in each offer. n - actual order in the experiment. R/F - correct answer of the question, FakeN - Fake representative offer created by NLG, EDG - Email delivered by Gmail, EDL - Email delivered by LinkedIn.}
\label{table-leg/fraud-ques}
\begin{tabular}{|c|c|c|c|c|}
\hline
Variable                                                                             & n  & R/F & Feature                                                              & DR \% (ACL) \\ \hline
\multirow{4}{*}{\begin{tabular}[c]{@{}c@{}}LinkedIn/Gmail \\ NLG/Human\end{tabular}} & 1  & F   & EDL/FakeN                                                            & 85.22 (3.76)      \\ \cline{2-5} 
                                                                                     & 3  & F   & EDG/Fake                                                             & 88.23 (3.7)       \\ \cline{2-5} 
                                                                                     & 5  & F   & EDL/Fake                                                             & 61.74 (3.47)      \\ \cline{2-5} 
                                                                                     & 7  & F   & EDG/FakeN                                                            & 94.11 (4.05)      \\ \hline
\multirow{4}{*}{Attack Type}                                                         & 2  & F   & \begin{tabular}[c]{@{}c@{}}EDG/Fake/\\ Sender\end{tabular}           & 64.70 (4.17)      \\ \cline{2-5} 
                                                                                     & 4  & F   & \begin{tabular}[c]{@{}c@{}}EDG/Fake/\\ Recipient\end{tabular}        & 70.58 (3.14)      \\ \cline{2-5} 
                                                                                     & 6  & F   & \begin{tabular}[c]{@{}c@{}}EDG/Fake/\\ Sender/Recipient\end{tabular} & 38.23 (3.64)      \\ \cline{2-5} 
                                                                                     & 8  & F   & EDG/Fake/None                                                        & 94.11 (4.14)      \\ \hline
\multirow{2}{*}{Legitimate}                                                          & 9  & F   & \begin{tabular}[c]{@{}c@{}}EDL/No Signal/\\ Fake Logo\end{tabular}    & 14.70 (3.97)      \\ \cline{2-5} 
                                                                                     & 10 & R   & EDG/No Signal                                                         & 41.17 (3.32)      \\ \hline
\end{tabular}
\end{table}

Participants' detection rate ranges from 4/10 (40\%) to 9/10 (90\%) (mean= 0.653, stddev=0.142, var=0.02). 
Average confidence level of participants ranges from 2.9 to 5 (mean=3.74, stddev=0.51, var=0.26). In order to have a better understanding of the effect of each variable, Table~\ref{part_one_overal_performance} provides overall performance in each category of representative offers. The last column in  Table \ref{table-leg/fraud-ques} and second column in Table \ref{part_one_overal_performance} is fraudulent Detection Rate which is the percentage of participants who detected the offer as fraudulent even if the offer itself is legitimate. This makes the comparison easier. So, for the \nth{10} offer and last two rows of Table \ref{part_one_overal_performance}, the actual performance of the participants is one minus the value in the cell (the value in the parentheses). In the case of Fraud WDR, last column of Table \ref{part_one_overal_performance}, SDR is 1 for offers detected as fake and -1 for offers detected as real {\em regardless} of whether the offer is real or fake.

The representative offer with fake logo has the least detection rate (14.7\%), and WDR shows that participants are highly confident that this is a legitimate offer. It shows that people do not pay attention to the logo of the company. Later, we check the participants' reasoning to see if they mentioned fake logo. Representative offer delivered by Gmail has the highest detection rate (91.7\%) and compared to the LinkedIn delivered ones, there is a gap of about 20\%. We shall see whether the difference is statistically significant.

\begin{table}[h]
	\centering
	\caption{Performance of participants on each category of offer (some offers are shared among two or more categories). n - number of offers in that category}
	\label{part_one_overal_performance}
	\begin{tabular}{|c|c|c|c|}
    \hline
    Category (n)       & Fraud. DR      & ACL  & Fraud. WDR   \\ \hline
    LinkedIn (2)       & 73.52         & 3.61 & 1.79         \\ \hline
    Gmail (2)          & 91.17         & 3.88 & 3.38         \\ \hline
    Human (2)          & 74.98         & 3.58 & 1.91         \\ \hline
    NLG (2)            & 89.7          & 3.91 & 3.26         \\ \hline
    Fake Offer (8) & 74.63         & 3.64 & 1.98         \\ \hline
    Real, No Signal (1)       & 41.17 (58.83) & 3.32 & 0.85 (-0.73) \\ \hline
    Real Job Offer (2) & 27.93 (72.07) & 3.51 & -1.11 (1.11) \\ \hline
    \end{tabular}
\end{table}


\subsection{Gmail and LinkedIn}
\label{subsec-gmail-linkedin}
The top two rows of Table~\ref{part_one_overal_performance} show the performance of participants on detecting LinkedIn and Gmail delivered representative offers. The results illustrate that success rate of deception attacks that are delivered by LinkedIn is higher than those delivered by Gmail. T-test shows a significant difference between performance of participants on LinkedIn and Gmail delivered representative offers (DR: p=0.009, df=53.183, t=-2.683, WDR: p=0.001, df=52.117, t=-3.283). To avoid multiple comparison problem, we utilized Benjamini-Hochberg~\cite{hochberg1990more} procedure to find the significant p-values. Both of them are still significant after applying the Benjamini-Hochberg procedure. So, there is a statistically significant difference between the effectiveness of attack using LinkedIn versus Gmail. 

We go one step deeper to see which personality types are more vulnerable to this kind of attack. Significant tests show that participants with lower \textit{Conscientiousness} level (p-value = 0.005) and lower \textit{Extraversion} level (p-value = 0.023) have significant difference in their detection rate for LinkedIn and Gmail delivered representative offers (p-value for participants with higher \textit{Openness} level is also less than 0.05 but Benjamini-Hochberg procedure made it non-significant).

\subsection{NLG and Human}
\label{subsec-nlg-human}
The average performance of participants on representative offers that are generated by NLG and representative offers that are generated by scammers (Human) can be found in the third and fourth rows of Table \ref{part_one_overal_performance}. The Detection rate for NLG is larger than the detection rate for Human, which shows that NLG generated emails are easier to detect.

Same as the Gmail and LinkedIn comparison, t-test (after applying Benjamini-Hochberg procedure) shows a significant difference between the performance of participants on NLG and Human generated representative offers (DR: p=0.01, df=63.233, t=-2.627, WDR: p=0.002, df=63.763, t=-3.197). This means our NLG generated offers are not as powerful as human-generated ones in fooling people. In Section \ref{sec-reasoning}, we go into details of participants' reasoning to have a better understanding of this difference.

\subsection{Real and Fake}
The average performance of participants on fake representative offers and the legitimate job offer (Real, No Signal) can be found in the fifth and sixth rows of Table~\ref{part_one_overal_performance}. The offer that we used for the fake logo is also a real job offer similar to \textit{Real, No Signal}. Since no one detected the fake logo, we also consider the representative offer with the fake logo as real. So, we have eight fake representative offers and two real ones. The last row of Table \ref{part_one_overal_performance} (Real) is the average performance for the real job offers (\textit{Real} is different from \textit{Real, No Signal}).

The detection rate of the Fake is larger than the detection rate for Real, which shows fraudulent emails are easier to detect, but the difference is not significant (p=0.076, df=46.731, t=-1.809).
The WDR is also higher for the fake offers which means participants are more confident about their answer when they choose Fraudulent than the Legitimate. In other words, participants tend not to choose Legitimate with high confidence, but again the difference is not significant (p=0.075, df=46.981, t=1.818).

\subsection{Attack Type}
\label{subsec-attack-type} 
We added two kinds of information to our emails, sender and receiver information. Different combinations of these variables result in four different emails. Table \ref{table-leg/fraud-ques} shows these four offers with participants' performance for each of them. Having both information in the email can dramatically reduce the detection rate (detection rate dropped from 94\% to 38\%). We conclude that emails that contain both sender and receiver information look more legitimate and are more capable of fooling people.

\subsection{Unmanipulated Variables} 
Unmanipulated variables are those that unlike independent variables are neither manipulated nor fixed in the experiment. So, these variables could affect the dependent variables. Here, we study the relation between these variables and dependent variables. We have different kinds of unmanipulated variables, some of them can be grouped together, e.g., age and sex are both part of participant demographics. Based on these similarities, we categorize the unmanipulated variables and study the correlation for each category separately. We have:
\begin{itemize}
	\item \textit{Personality Traits}: Five different traits defined by Big Five Personality Test
	\item \textit{Demographics}: Age, gender
	\item \textit{Knowledge and Background}: Participants' background knowledge on computer and email, and their education level. Emails sent each day, Email propagation knowledge, Social network usage and education level are in this category.
\end{itemize}

\subsubsection{Personality Traits}
\label{subsubsec-personality} 
Personality traits reveal internal aspects of each person's mind. 
Since the performance of participants and the personality trait scores are not continuous variables, it is better not to use the Pearson correlation test between each trait and the performance value directly. Instead of using correlation test directly, we calculate the correlation between participants' ranks on DR, WDR, and ACL and their ranks on each of the personality traits, but none of the correlations are significant.

We also group the participants into two groups based on each personality trait and perform a t-test for comparing them. For each trait, suppose M is the median. We divide the participants into two groups based on the value of M; those who are greater than M and those who are less than or equal to M. Then we apply t-test to compare their performance.
Table~\ref{table-sig-personality} presents p-values of t-test on different groups of participants. The t-test shows no significant difference between participants' personality and their performance and confidence level.

\begin{table}[h]
	\centering
	\caption{P-value of t-test for comparing the performance of participants based on each trait}
	\label{table-sig-personality}
	\begin{tabular}{|c|c|c|c|}
\hline
Trait (median)         & DR    & WDR   & ACL   \\ \hline
Extraversion (27)      & 1     & 0.995 & 0.846 \\ \hline
Agreeableness (33)     & 0.129 & 0.991 & 0.667 \\ \hline
Conscientiousness (32) & 0.989 & 0.242 & 0.209 \\ \hline
Neuroticism (23)       & 0.73  & 0.982 & 0.348 \\ \hline
Openness (35)          & 0.662 & 0.452 & 0.227 \\ \hline
\end{tabular}
\end{table}

\subsubsection{Demographics}
\label{subsubsec-demography}
Demographic features of participants can also have an effect on their responses. We check the significance of the difference between mean values of dependent variables as a function of age and sex. For the \textit{age}, we divided the participants into two groups: older than 21 years old (median) and younger than 21. Based on the t-test result, there is no statistically significant difference in participants' average performance with different age and sex.

\subsubsection{Knowledge and Background}
Users' background knowledge and experience in working with email and other applications like social networks may have an effect on their performance. Hence, we test the correlation between these variables (email propagation knowledge, email sent each day, social network usage, education level, and years using email) and performance indicators. Applying correlation test on them does not show any significant correlation between them. 

\subsection{Strategy Analysis}
\label{subsec-signal-analysis}
In each representative offer of the legitimate/fraudulent section, besides asking ``Fraudulent/Legitimate?'', we also asked participants to write down their entire reasoning for their decision. The reasons used by participants to distinguish between fraudulent and legitimate offers is another important aspect of this study. Knowledge of these strategies can be used by defenders to devise a new generation of attacks and test their filters, and also to identify the things that email users do not pay attention to and need help with.

\begin{table}[t]
	\centering
	\caption{Number of participants using each strategy in the first section. N - No. of participants, PG - performance gain}
	\label{table-masq-strategy}
\begin{tabular}{|c|c|c|c|}
\hline
Strategy                              & N  & \%   & PG     \\ \hline
Asking for Action/Info                & 32 & 94.1 & -0.209 \\ \hline
Legality Claim                        & 26 & 76.4 & 0.020  \\ \hline
Phishing Hints                        & 24 & 70.5 & 0.018  \\ \hline
Sender Info                           & 21 & 61.7 & -0.038 \\ \hline
Job Info                              & 20 & 58.8 & -0.080 \\ \hline
Grammar, Capitalization, Punctuation & 19 & 55.8 & 0      \\ \hline
LinkedIn                              & 13 & 38.2 & -0.073 \\ \hline
Sender's Email                        & 12 & 35.2 & 0.008  \\ \hline
Receiver Info                         & 10 & 29.4 & -0.103 \\ \hline
Spelling Issue                        & 10 & 29.4 & -0.046 \\ \hline
Profile Picture                       & 9  & 26.4 & -0.072 \\ \hline
Practical Real World Consideration                                  & 9  & 26.4 & 0.033  \\ \hline
Company Info                          & 8  & 23.5 & -0.036 \\ \hline
Email Production and Delivery                                   & 7  & 20.5 & 0.077  \\ \hline
Over-thinker                          & 4  & 11.7 & -0.088 \\ \hline
Fake Logo                             & 0  & 0    & 0      \\ \hline
\end{tabular}
\end{table}

We categorized participants' strategies and the signals that we added to the offers (e.g. fake logo and LinkedIn) into 16 groups as follows (the steps that we took to extract these strategies are recommended by \cite{eaves2001synthesis}):

\begin{itemize} 
	\item \textbf{Asking for action/info:} Does the email ask for doing an action? or ask for any information?
	\item \textbf{Legality Claim:} The email claims that the offer is legal
	\item \textbf{Phishing Hints:} The email tries to deceive the reader or talks about doing payment 
	\item \textbf{Sender Info:} Sender's contact info (address, phone, and company name) are provided or not
	\item \textbf{Job Info:} Enough details about the job are provided or not
	\item \textbf{Grammar}, \textbf{capitalization}, and \textbf{punctuation} 
	\item \textbf{LinkedIn:} The message was delivered through LinkedIn, not through Gmail 
	\item \textbf{Sender's Email:} All the aspects related to the sender's email address, e.g., using public email servers, emails with different top-level domain than the actual company (info@chase.opportunity.com), and email whose domain is same as the company name (info@microsoft.com)
	\item \textbf{Receiver's Info:} Sender has some information about the receiver (his name or educational background)
	\item \textbf{Spelling Issue:} Misspelled words in the Email 
	\item \textbf{Profile Picture:} Anything suspicious or non-suspicious about the LinkedIn profile picture, e.g., ``profile pic is a white girl while the company is from India'', ``unprofessional profile picture'', etc.
	\item \textbf{Practical Real World Consideration:} The email asks to reply to another email or requests for further contact to share more information
	\item \textbf{Company Info:} Enough details about the company are provided or not
	\item \textbf{Email Production and Delivery:} All the details about the structure, writing and author of the emails, e.g., spacing error, poor formating, CEO sending email, and using letterhead	
	\item \textbf{Over-thinkers:} Participants who argued that spelling and grammar issues show that the email is real, since if it was from an attacker, it would not have such mistakes	
	\item \textbf{Fake Logo:} Fake LinkedIn logo is used
\end{itemize}

Table~\ref{table-masq-strategy} shows how many participants used each type of strategy. \textit{Asking for action/info}, \textit{legality claim} and \textit{phishing hints} are the most used strategies. To have a better understanding of the difference between strategies, we define a metric called \textit{Performance Gain} (PG). The idea is similar to information gain, but for PG, we calculate the difference between the average performance of participants who used a specific strategy $s$ and the average performance of those who did not use $s$ ($PG(s) = AvgDR_{used}(s) - AvgDR_{notused}(s)$). Performance gain for all of the strategies is very low. This shows that paying attention to a clue itself without knowing how to leverage that clue is not enough to correctly identify the phishing/legitimate emails. There are also some negative performance gains which mean there are some strategies that fooled people instead of helping them to detect the phishing emails. We cannot generalize our conclusion since we have more than one clue in each representative offer and the priority of strategies differs for each participant. 

These strategies along with their importance (priority) for users' decision-making process can be used to improve the effectiveness of detectors by subjecting them to new, more effective attacks, or to train people about phishing emails.

Studying the effect of strategies used by participants on their performance is not straightforward. To achieve this goal we need to find a way to group participants based on their strategies (similarity of strategies that they used). Once we group them together, we can compare the performance of users in each group.

We present two different methods for grouping participants together (a third based on the number of identified signals yielded nothing significant). In the first method, we use the 16 extracted strategies as boolean features for participants and clustered participants based on the feature vectors. Since we do not know the ideal number of clusters to use, hierarchical agglomerative clustering \cite{murtagh1983survey} is used. We tried both cosine similarity and Euclidean distance as distance metrics. The results with Euclidean distance were more reasonable than with cosine similarity, so we report the ones for  Euclidean distance. The output of hierarchical clustering is a tree whose leaves are the individual feature vectors. By cutting the tree at every height we get different clustering output, some of them may have a singleton cluster. We try different cutting points and keep only those clusterings that do not have any singleton cluster for further analysis. 
In the second method of grouping, we use \textit{sophistication level} to separate participants, where  \textit{sophistication level} is the total number of {\em groups} of strategies that a participant used. 

\begin{table}[t]
\centering
\caption{Average DR, WDR, and ACL for each group. Group: ``cluster number'' in \textit{Clustering} and ``number of strategy groups'' in \textit{Sophistication Level}. N - number of participants}
\label{table-performance-gourping}
\begin{tabular}{|c|c|c|c|c|c|}
\hline
\begin{tabular}[c]{@{}c@{}}Grouping \\ Method\end{tabular}                       & Group         & N  & Average DR & Average WDR & ACL  \\ \hline
\multirow{3}{*}{Clustering}                                                      & 0             & 11 & 0.71       & 1.78        & 3.55 \\
                                                                                 & 1             & 4  & 0.47       & 0.62        & 3.8  \\
                                                                                 & 2             & 19 & 0.65       & 1.26        & 3.83 \\ \hline
\multirow{6}{*}{\begin{tabular}[c]{@{}c@{}}Sophistication\\  Level\end{tabular}} & \textless5    & 4  & 0.65       & 1.9         & 4.15 \\
                                                                                 & 5             & 6  & 0.76       & 1.69        & 3.91 \\
                                                                                 & 6             & 8  & 0.67       & 0.98        & 3.46 \\
                                                                                 & 7             & 5  & 0.58       & 0.97        & 3.56 \\
                                                                                 & 8             & 7  & 0.67       & 1.74        & 3.7  \\
                                                                                 & \textgreater8 & 4  & 0.5        & 0.87        & 3.92 \\ \hline
\end{tabular}
\end{table}

\begin{figure}[h]
	\centering    
	\includegraphics[width=3.2in]{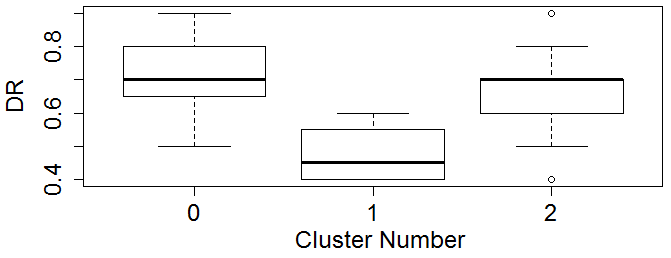}	
	\caption{Clusterwise participants’ DR distribution}
	\label{fig-boxplot-masq-DRcluster}
\end{figure}

\begin{figure}[h]
	\centering    
	\includegraphics[width=3.2in]{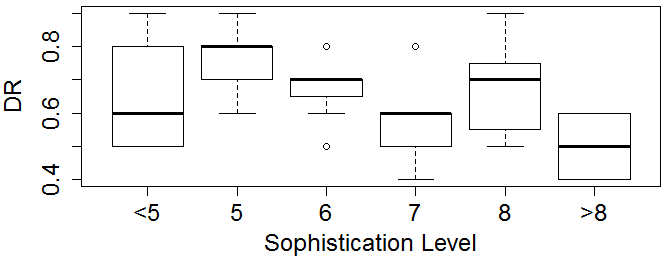}
	\caption{Participants’ DR distribution on different sophistication level}
	\label{fig-boxplot-masq-DRsophi}
\end{figure}

Table~\ref{table-performance-gourping} shows average performance of participants for aforementioned grouping methods. Distribution of the detection rates is also depicted in Figures \ref{fig-boxplot-masq-DRcluster} and \ref{fig-boxplot-masq-DRsophi}. For the \textit{Clustering} method, we used one of the clustering outputs that has three classes {\em as an example here}. The results for other cutting points are similar to this one. In the \textit{Clustering} method, cluster zero does perform better than two other clusters. We check later if the difference is statistically significant or not. In the \textit{Sophistication Level}, there is a rise and then almost a straight drop in performance with one exceptional group. Among these levels, sophistication level five is the best performing category (DR=0.76), but if we consider confidence level as well, the minimum sophistication level (\textless5) is the best performing (WDR=1.9). The top performer used the following strategies \textit{Sender Info}, \textit{Asking for action/info}, \textit{Grammar, capitalization, and punctuation}, \textit{Legality claim}, and \textit{Practical Real World Consideration}.

We apply ANOVA tests on the two aforementioned grouping methods to see whether or not the performance of each group differs statistically significantly. Table~\ref{table-anova} shows the p-value of ANOVA tests on different grouping method. All the p-values are bigger than 0.05 except for the detection rate of the hierarchical clustering method (F(2, 31) = 5.26, p-value = 0.01).\footnote{The p-value of 0.06 for Sophistication is also close to our 0.05 threshold} A Tukey post-hoc test reveals that the cluster number zero is significantly different from two other clusters. Participants in this cluster do not pay attention to \textit{Sender} and \textit{Receiver} info. This suggests that there are some strategies that can protect people better from deception attacks. However, we hesitate to generalize this conclusion since there are some features that we did not consider, e.g., IQ, native English speaker, etc. 

\begin{table}[h]
	\centering
	\caption{P (F) value of ANOVA tests on different grouping methods}
	\label{table-anova}
	\begin{tabular}{|c|c|c|c|}
		\hline
		Method         & DR                   & WDR          & ACL         \\ \hline
		Clustering     & \textbf{0.01 (5.26)} & 0.148 (2.03) & 0.35 (1.06) \\ \hline
		Sophistication & 0.06 (2.37)          & 0.44 (0.98)  & 0.25 (1.4)  \\ \hline
	\end{tabular}
\end{table}

Different people pay attention to a subset of all fake signals that exist in the email. An interesting question here is ``Do people with different characteristics have different strategies?'', e.g., ``Do people with higher \textit{conscientiousness} use more signals that are less important for those with lower \textit{conscientiousness}?'' or ``Do women use set of signals that are less important for men?'' Here, we choose the number of strategies as a response variable and test the effect of other variables on this response variable.

Comparing the number of strategies used by males and females reveals that males use more strategies than females (7.2 versus 5.8) and the difference is significant with ninetyfive percent confidence level (p=0.024, df=31.25, t=2.364). Besides that, there are also some differences in the type of strategies used by male and females. For example, none of the females used \textit{Email Production and Delivery} as their strategy. Also, females pay more attention to the sender information and spelling errors than males.


The number of strategies used by participants may vary for people with different personality. To study the relation between the number of strategies and the personality traits, again we divide the participants into two groups based on their traits (higher and lower than median) and compare the strategies used by them. We did this for all five traits, but no significant difference was found between strategies used by different groups.

\subsection{Linear Regression Model} 
\label{subsec-linear-reg}
A linear regression model is another way of analyzing the relation between the variables. First, we create a model for the detection rate considering one predictor at a time. Predictor variables are all unmanipulated and independent variables in our study. We also consider the \textit{number of strategies used by each participant} (\textit{Strategy Count}) as a variable since it can be interpreted as a sophistication level of participants. 

In all generated models, the p-values for predictors is bigger than 0.05 except for the \textit{Strategy Count} (p-value = 0.015, $R^2$ = 0.171). Since there are several unpredictive variables in the model, we should only keep those variables that have a minimum contribution to the model. So, we use 0.04 as a threshold for the $R^2$ value and only consider variables for which the corresponding model has the $R^2$ value bigger than 0.04. Gender, age, education level, and \textit{Strategy Count} are the variables that passed our $R^2$ criteria.\footnote{For $m = 4$ independent variables, we need at least $4(m+2) = 24$ independent observations, m coefficients + intercept + variance is $m+2$, and we have $34 > 24$ so we can construct a regression model.} We created the model based on these variables, and Table \ref{table-linear-model} displays the coefficients, standard error, \textit{t} value, and \textit{p} values for each of the fixed effects. The $R^2$ is 0.268 (adjusted $R^2$ = 0.105) and \textit{Strategy Count} is the only statistically significant predictor. The coefficient for the \textit{Strategy Count} is negative, which means employing too many strategy groups led to a lower performance for our participants. We repeated this analysis for WDR and ACL as response variables but none of the results are significant.

\begin{table}[h]
	\centering
	\caption{Multiple regression model for predicting detection rate}
	\label{table-linear-model}
	\begin{tabular}{lcccc}
		\hline
		Fixed Effect        & Coefficient & Std Error & \textit{t}      & \textit{p}               \\ \hline
		(Intercept)         & 0.636       & 0.168     & 3.770  & \textless 0.001 \\
		Age                 & 0.012       & 0.008     & 1.515  & 0.141           \\
		Gender (Male)       & -0.015      & 0.055     & -0.278 & 0.782           \\
		Education (Diploma) & -0.419      & 0.349     & -1.199 & 0.241           \\
		Education (Master)  & -0.121      & 0.110     & -1.103 & 0.279           \\
		Education (PhD)     & -0.079      & 0.089     & -0.892 & 0.380           \\
		Strategy Count      & -0.034      & 0.014     & -2.340 & \textless 0.05  \\ \hline
	\end{tabular}
\end{table}

\section{Reasoning Section}
\label{sec-reasoning}
The second section of the experiment is about the reasoning of our participants on nine representative offers involving fake name, fake logos, NLG generated emails and customization. Customization is done based on participants' education and/or work background. Table~\ref{scaff-questions-part2} shows details of the offers in this section (\nth{6} is a company representative fraud email that we used without any change).
Recall that, for this section, we told participants that all emails are fraudulent and they need to find all indicators that show the emails are fake.

Most of the strategies are similar to those in Section \ref{subsec-signal-analysis}. Table \ref{table-masq-strategy_2} shows all the strategies used by participants. There are two new strategies:
\begin{itemize}
	\item \textbf{Fake Company name:} Did participants detect the fake company name for the representative offer from the \textit{Donald Duck and Mickey Mouse} company?
	\item \textbf{Constructed:} Email created by the human or by the computer (using NLG tech.).
\end{itemize}

Seven participants noticed that some emails are generated by the computer. A deficiency in the grammar that we used for NLG can be the reason for this observation. This can also justify the higher detection rate of NLG generated emails compared to human-generated emails in the previous section of the experiment (legitimate/fraudulent). Also, some of the participants used \textit{LinkedIn} as their strategy in the first section of the experiment, but in this part, since we told them that all the emails are fraudulent, no one mentioned it.

\begin{table}[h]
	\centering
	\caption{Nine representative offers in the reasoning section. n - actual order in the experiment, FakeN - Fake representative offer created by NLG, EDG - Email delivered by Gmail, EDL - Email delivered by LinkedIn.}
	\label{scaff-questions-part2}
	\begin{tabular}{|c|c|c|}
\hline
Variable                    & n & Feature                                                                                  \\ \hline
\multirow{3}{*}{Customized} & 1 & EDL/FakeN/Custom\_Work\_Background                                                       \\ \cline{2-3} 
                            & 3 & EDL/FakeN/Custom\_Education\_Background                                                  \\ \cline{2-3} 
                            & 8 & \begin{tabular}[c]{@{}c@{}}EDL/FakeN/\\ Custom\_Education\&Work\_Background\end{tabular} \\ \hline
\multirow{4}{*}{Fake Logo}  & 2 & EDL/FakeN/Fake\_Logo\_1                                                                  \\ \cline{2-3} 
                            & 5 & EDL/FakeN/Fake\_Logo\_2                                                                  \\ \cline{2-3} 
                            & 7 & EDL/Fake/Fake\_Logo\_3                                                                   \\ \cline{2-3} 
                            & 9 & EDL/FakeN/Fake\_Logo\_4                                                                  \\ \hline
Company Name                & 4 & EDL/FakeN/Fake Company                                                                   \\ \hline
N/A                         & 6 & EDG/Fake                                                                                 \\ \hline
\end{tabular}
\end{table}

\begin{table}[h]
	\centering
	\caption{Number of participants (N) using each strategy in the second section.}
	\label{table-masq-strategy_2}
	\begin{tabular}{|c|c|c|c|}
\hline
Strategy          & N  & Strategy                                                                        & N  \\ \hline
Legality Claim    & 27 & \begin{tabular}[c]{@{}c@{}}Grammar, Capitalization, \\ Punctuation\end{tabular} & 25 \\ \hline
Fake Company Name & 22 & Email Production and Delivery                                                   & 19 \\ \hline
Sender's Email    & 19 & Asking for Action/Info                                                          & 19 \\ \hline
Company Info      & 19 & Phishing Hints                                                                  & 18 \\ \hline
Sender Info       & 17 & \begin{tabular}[c]{@{}c@{}}Practical Real World \\ Consideration\end{tabular}   & 11 \\ \hline
Spelling Issue    & 10 & Receiver Info                                                                   & 9  \\ \hline
Job Info          & 9  & Profile Picture                                                                 & 8  \\ \hline
Constructed       & 7  & Fake logo                                                                       & 3  \\ \hline
Over-thinker      & 2  &                                                                                 &    \\ \hline
\end{tabular}
\end{table}

Only three participants detected the fake logos. Two of them detected only the easiest one and the third participant detected the first two easy ones. This shows that participants cannot easily tell the difference between visual cues, so one cannot rely on this kind of authentication (e.g. banks and financial companies have started using images) and phishers can escape detection even if their images are not exactly similar. Even though most of the phishing emails use genuine logos~\cite{blythe2011f}, but this can help the attacker to avoid detection techniques that rely on image similarity as a feature.

Same as in the Legitimate/Fraudulent part, \textit{Legality Claim} has a very high frequency. This shows that most of the participants know that when something is legal, there is no need to mention its legality explicitly.

Some participants were suspicious of the location of the company, e.g. being from India, China, etc. (considered as \textit{Company Info}). The relation between the profile picture of the sender (race) and the location of the company (\textit{Profile Picture}) and the relation between the name of the company and their expertise (\textit{Company Info}) are two other interesting points brought up by participants. 

In the first section of the experiment, we asked participants to decide whether the representative offer is real or fake, and, we asked them to write down all of their reasoning. It is possible that participants ignore the rest of the email after they have made their decision. In the second section, we told them that all the emails are fraudulent and they just need to say why these are fake. Comparing the average number of strategies per question in the first and second sections can help us to see if this really happened or not. The average number of strategies used in each question in section two (1.78) is higher than section one (1.46) and the difference is significant (p=0.01, df=15.581, t=-2.92).
So, it is possible that some participants did not pay attention when we told them to write down \textbf{all} their strategies. However, note that the content and design of the emails was not exactly the same for the two parts. So, we cannot be certain about this. 

We also have three customized representative offer in this part. The goal is to see if the customization can affect the reasoning of the participants or not. The average number of strategies in each question can be used as a signal to study the changes in the reasoning of participants. The average number of strategies for customized offers (1.62) is a little lower than for uncustomized ones (1.86), the difference is although not significant, but close (p-value = 0.06). 

\section{Interview}
\label{sec-interview}
At the end of the experiment, we did a short interview (about 10 minutes) with the participants. The interview had a combination of some  questions that required verbal answers, e.g., ``What is the difference between CC and BCC in sending email?'', and some that required demonstrations, e.g., ``Can you show us how you can see full email header?'' Table \ref{table-interview-questions} shows all the questions that we asked during the interview. There are two types of questions, simple and interactive. simple ones are those that we just ask a question and the participants give answers to it. The interactive question, on the other hand, has two parts. First part is same as the simple one, a question that interviewer asks from participants. But the participants need to look at another content (an email for example) to answer the question. For example, in the fifth question, instead of asking them ``Do you know what is the email full header?'', we ask them to log into their email client, and then show us the full header of an email. Someone may have heard about the full header but still may not know how to check it. This helps us to differentiate between them. Also in the sixth question, we show participants an email that someone shares the report of a project via a Dropbox link (we asked them to suppose the sender is their colleague that work with them in a same project), but the actual link is pointing to a fake website (in the html \textless a\textgreater~tag, href attribute is different from the tag text). Here the goal is to check participants' knowledge about this technique without asking them directly.

\begin{table}[h]
\centering
\caption{Interview questions}
\label{table-interview-questions}
\begin{tabular}{|c|l|c|}
\hline
n & \multicolumn{1}{c|}{Question}                                                                                                                                                         & Type        \\ \hline
1 & \begin{tabular}[c]{@{}l@{}}What is the difference between CC and BCC \\ in sending email?\end{tabular}                                                                                & Verbal      \\ \hline
2 & \begin{tabular}[c]{@{}l@{}}How many parts there are in the Email? (You\\  always see content, what about other parts?)\end{tabular}                                                   & Verbal      \\ \hline
3 & How Emails are sent through the Internet?                                                                                                                                             & Verbal     \\ \hline
4 & \begin{tabular}[c]{@{}l@{}}If you think an Email is fake, do you check \\ its header? Which fields?\end{tabular}                                                                      & Verbal      \\ \hline
5 & \begin{tabular}[c]{@{}l@{}}Can you show us how you can see full email\\  header?\end{tabular}                                                                                         & Demonstrative \\ \hline
6 & \begin{tabular}[c]{@{}l@{}}Can you find out why this is a suspicious Email?\\ (an email has been shown to the participant)\end{tabular}                                               & Demonstrative \\ \hline
7 & \begin{tabular}[c]{@{}l@{}}When you are going to detect if an Email is real\\  or fake how much do you rely on: Sender email \\ address, date/time, and domain of sender\end{tabular} & Verbal      \\ \hline
\end{tabular}
\end{table}

20 participants know the difference between CC and BCC, eight only know CC and six do not know any of them. There is no significant difference between detection rate of those participants who know about the BCC and the rest of the participants. Knowledge about the email transmission (sender server, receiver server, and relays) is too technical, and as we expected, only two participants had this knowledge. Both of them are computer science graduate students.

For the full email header, 25 participants know how to just check the simple fields of the header (from, to, subject, and date), three participants do not know anything about the header and six participants have the knowledge to show the full header (we showed them the user interface of their own email client to remove the effect of the different user interface). Here also all of these six participants are computer science students (five graduate and one bachelor). However, no significant difference was found between detection rate of those six participants and the rest of participants (p-value = 0.833). In the sixth question, only five participants recognized that the actual link is different from what has been shown (no significant difference between their detection rate, p-value = 0.123). 

The last question which is a multiple choice question consists of three subparts. All three subparts are about the situation that they receive an email and they want to decide if it is fake or real. We asked them how much they rely on the \textit{sender's email address}, or the \textit{domain of the sender}, or the \textit{date/time}. In order to make it clear for everyone, during the interview, the interviewer mentioned to all the participants that by \textit{domain of the sender} we mean if the email is from general email services like @gmail.com, @outlook.com, etc or it is from companies private domain, e.g. @google.com, @nyu.edu, etc.  For each of the subparts, participants are given three choices ``a lot'', ``a little'' and ``not at all.'' Only one participant answered that he/she does not rely on \textit{sender email} at all, 23 participants answered ``a lot'' and 10 participants answered ``a little''. We checked the reasons given in the first section of the experiment by the person who answered ``not at all'', and he/she did not list sender email as their strategy. Surprisingly, only eight participants out of 23 that chose ``a lot'' explicitly listed sender email as their strategy in the first section of the experiment (and four out of 10 who chose ``a little''). This is interesting that only 36\% of participants that said they pay attention to the sender email address explicitly mentioned it as one of their strategies in the first section of the experiment. 
We also compared the detection rate of the participants who answered ``a lot'' and the rest of participants but the difference is not significant (p-value = 0.085).

For the \textit{domain of the sender}, 21 participants chose ``a lot'', nine chose ``a little'' and four participants chose ``not at all''. Again, the important thing is, do they really explicitly mention the domain of the sender in the first section of the experiment? In the experiment, we had some emails from LinkedIn and some from other free email services. Same as for ``sender email'', a few participants who answered ``a lot'' or ``a little'' to relying on ``sender domain'' mentioned LinkedIn in their reasoning, nine out of 21, three out of nine. Also, one of the participants out of four whose answers were ``not at all,'' referred to LinkedIn in his/her reasoning. However, we could not find any significant difference in their performance (p-value = 0.228).

Participants answer to \textit{date/time} question is quite different from two previous subparts. Only two participants answered ``a lot'', 20 ``a little', and 12 ``not at all''. So, most of the participants think that date/time does not have that much effect on their decision compared to \textit{sender email} and \textit{domain of the sender}. Unfortunately, we did not have any signal related to the date/time to check if participants pick on this signal.

\section{Limitations}
This study did not consider several variables in its design and analysis.

We used the job scams directly from the existing datasets without any modifications. So, we did not control the clues that exist in each offer. For example, one offer can have grammar/writing issues and another one can be grammatically correct. This affects participants' decision and as a result  the average detection rate of participants in each study group.

In the first section of the study, we used two legitimate offers and eight fake offers. This can bias participants decision into not labeling offers as fake since they might feel they labeled everything as fake. Adjusting the ratio of classes to not bias the participants is challenging in a study like ours.

We showed eight fake emails with three different variables (LinkedIn/Gmail, NLG/Human, and four attack types) to participants. This might raise the concern about the learning effect of showing so many variables to each participant. Using a fixed ordering for the offers deteriorates the issue. An improvement to this study would randomize the offers presented to the participants.

When we checked participants' reasoning, we realized some of the changes that we made to the fake offers created some unexpected clues. For example, for one of the offers, participants noticed a gender mismatch between the name of the sender and his/her profile picture. An overall improvement in consistency and credibility of the fake offers would strengthen the study.

\section{Related Work}
Research on phishing and social media can be divided into three categories: 1) attacks on social media \cite{bilgeSBK09,mahmood2012your,huber2011friend}, 2) human interaction with phishing/spam emails and social networks \cite{tang2011s,meijdam2015phishing,karavaraslow}, and 3) detection and authentication mechanism \cite{aassalBDV20,dasBAV20,egoziV18,herzberg,oppligerG05,verma2019,zhouV20}.

\textbf{Attacks on Social Networks}. These attacks are variants of traditional
security threats that leverage social networks as a new medium. Identity theft attack \cite{bilgeSBK09,huber2011friend,huber2010earth} and reverse social engineering attack \cite{irani2011reverse} are two types of these attacks. 
The social scaffolding attack (using LinkedIn) that we introduced in this study is a new type of attack on social networks. Previous studies have analyzed the effect of using Facebook as a medium for delivering the attack messages \cite{benenson2014susceptibility, benenson2017unpacking} but their finding contradicts each other. Researchers in \cite{benenson2014susceptibility} found fraudulent messages delivery by email more successful in fooling users to click on a link embedded in the message, while authors in \cite{benenson2017unpacking} found messages delivered by Facebook more convincing.

\textbf{Defense Mechanism}.
System- and user-oriented are two categories of research related to the defense mechanism. 
The system-oriented approach tries to create different computer models to stop the existing attacks, e.g. detecting Identity Clone attacks \cite{jin2011towards,he2014defence,yang2014uncovering} or privacy control. On the other hand, the user-oriented approach tries to increase user awareness about fraudulent activities. There is a good survey on different factors that have an impact on individual susceptibility to malicious online influence \cite{williams2017individual}. Understanding these factors can help to effectively reduce potential vulnerabilities and develop more effective and targeted mitigations.

\textbf{User Study}. 
Understanding the relationship between users' characteristics and their behaviour when they encounter phishing emails/websites has been done previously \cite{dhamija2006phishing,sheng2010falls,downs2006decision,vishwanath2015habitual,egelman2008you,karakasiliotis2006assessing,neupane2015multi}. They also analyzed the users' decision making process when they decide to respond to an email or not. 
The relation between demographic features and phishing susceptibility indicates that women and participants between 18 to 25 are more susceptible than other participants \cite{sheng2010falls}. Authors in \cite{karakasiliotis2006assessing} show that in the case of message content alone, many users face a hard task to differentiate between a genuine email and a bogus one. Their focus is on content, not mimicking the scaffolding to deceive. Researchers in \cite{blythe2011f} show that phishing emails with logos are more capable of fooling people. They did their study with real phishing emails, which have original logos of the companies, but they did not do any further investigation to see if there is any difference between the effectiveness of the attack using fake or original logos. We showed that people do not pay attention to the originality of the logos.

Analysis of users' behavior can be done by using the strategies that they used. Authors in \cite{downs2006decision} showed that even though people are good at managing the risks that that they are aware of, they do not perform well in cases of unfamiliar risks. We can conclude that scaffolding can be successful since most of the people believe that messages in a professional social network like LinkedIn are legitimate. Our experiment is similar to this work but they created their own phishing emails by putting some fake signals together, whereas we used a combination of existing company representative fraud email and crafted ones. Since they had a set of specific clues for each email, they could not analysis participants' \textit{strategy set}s and their relation to the performance of detecting fake emails.
The relation between habitual use of social network and probability of getting deceived was also studied in \cite{vishwanath2015habitual}. Their study shows that both habitual and ``inhabitual'' users accept the friend requests, but habitual users are more susceptible to providing information to the phisher. Interestingly, there is some research that shows Facebook does not seem to have the same effect as LinkedIn. Researchers in \cite{benenson2014susceptibility} find that users were more likely to click on a link in email versus in a Facebook message. LinkedIn and Facebook have different goals, which may explain the difference. Researchers in \cite{alqarni2016toward} used demographics, anonymity, social capital and risk perception of Facebook users to predict their susceptibility to phishing attacks. They used questionnaire approach to test their hypotheses which is not an accurate method based on our findings that people do not have an explicit knowledge of their strategies. In other words, either people are not aware of the strategies they use, or they do not explicitly mention some of them.



Although LinkedIn has been used as a mechanism for spreading phishing emails \cite{silic2016dark}, none of the previous works studied the scaffolding separately as a mechanism for improving the effectiveness of a deception attack. Besides this, we also, analyze the strategies that participants used and the correlation between the demographics and their performance. Also, the effectiveness of natural language generation in email masquerade attack has been studied in \cite{baki2017scaling}, but here we use it for generating company representative offers. We believe that NLG may not have worked as well for offers since they tend to be longer and more complex. 

\section{Conclusions and Future Work}
We conducted a study in which we parameterized a deception attack and observed how users deal with it. We built our study based on the company representative fraud. Using LinkedIn instead of traditional email-based attacks, different level of information about sender and receiver, and using fake logos are some of the parameters that we studied.

Our study showed that putting a simple deception attack into the context of LinkedIn can increase the attack success rate significantly. People trust messages received from a professional network like LinkedIn more than normal emails. We also analyzed the participants' strategies among other variables. Our analyses show that the participants who use fewer strategies perform better than the other participants. We also found that the strategies used by females and males are different from each other. Males use more strategies than females, and they pay more attention to the formatting of the emails. Yet, interestingly, the performance is not significantly different between the two groups.

Adding sender's contact information and also making the email customized for each particular receiver, by adding receiver's name in the greeting, can make the message seem more legitimate. We also used natural language generation (NLG) technique to semi-automatically generate the attack, but this attack was not so successful, and participants performed better in detecting NLG generated emails.

The interview that we did with participants at the end of the experiment revealed that people may not use or explicitly mention the strategies that they use. Only 36\% of the participants who said they pay attention to the sender's email address a lot, actually mentioned the sender email in the first section of the experiment. Our findings suggest that training regimes need to: reinforce that less is more, focus on a few key strategies, and emphasize that a professional network is not necessarily more trustworthy. Attackers can infiltrate any network.

As with all new works,  this represents an initial attempt to explore the space of social trust exploitation, and even though it has the above limitations, we believe it could help other researchers. We plan to improve the work in the future so that the experimental setup addresses the limitations.

\section*{Acknowledgments}
This research was partially supported by NSF grants CNS 1319212, DUE 1241772, DGE 1433817 and CNS 1527364. It was also partly supported by  U. S. Army Research
Laboratory and the U. S. Army Research Office under
contract/grant number W911NF-16-1-0422. 




\bibliographystyle{IEEEtran}
\bibliography{IEEEabrv,main}

\begin{thebibliography}{10}
\providecommand{\url}[1]{#1}
\csname url@samestyle\endcsname
\providecommand{\newblock}{\relax}
\providecommand{\bibinfo}[2]{#2}
\providecommand{\BIBentrySTDinterwordspacing}{\spaceskip=0pt\relax}
\providecommand{\BIBentryALTinterwordstretchfactor}{4}
\providecommand{\BIBentryALTinterwordspacing}{\spaceskip=\fontdimen2\font plus
\BIBentryALTinterwordstretchfactor\fontdimen3\font minus
  \fontdimen4\font\relax}
\providecommand{\BIBforeignlanguage}[2]{{%
\expandafter\ifx\csname l@#1\endcsname\relax
\typeout{** WARNING: IEEEtran.bst: No hyphenation pattern has been}%
\typeout{** loaded for the language `#1'. Using the pattern for}%
\typeout{** the default language instead.}%
\else
\language=\csname l@#1\endcsname
\fi
#2}}
\providecommand{\BIBdecl}{\relax}
\BIBdecl

\bibitem{kelleyB16}
T.~Kelley and B.~I. Bertenthal, ``Real-world decision making: Logging into
  secure vs. insecure websites,'' in \emph{USEC}, 2016.

\bibitem{schechter2007emperor}
S.~E. Schechter, R.~Dhamija, A.~Ozment, and I.~Fischer, ``The emperor's new
  security indicators,'' in \emph{2007 IEEE Symposium on Security and Privacy
  (SP'07)}.\hskip 1em plus 0.5em minus 0.4em\relax IEEE, 2007, pp. 51--65.

\bibitem{alsharnouby2015phishing}
M.~Alsharnouby, F.~Alaca, and S.~Chiasson, ``Why phishing still works: user
  strategies for combating phishing attacks,'' \emph{International Journal of
  Human-Computer Studies}, vol.~82, pp. 69--82, 2015.

\bibitem{wangHCVR12}
J.~Wang, T.~Herath, R.~Chen, A.~Vishwanath, and H.~R. Rao, ``Phishing
  susceptibility: An investigation into the processing of a targeted spear
  phishing email,'' \emph{{IEEE} Trans. Prof. Communication}, vol.~55, no.~4,
  pp. 345--362, 2012.

\bibitem{neupane2015multi}
A.~Neupane, M.~L. Rahman, N.~Saxena, and L.~Hirshfield, ``A multi-modal
  neuro-physiological study of phishing detection and malware warnings,'' in
  \emph{Proceedings of the 22nd ACM SIGSAC Conference on Computer and
  Communications Security}.\hskip 1em plus 0.5em minus 0.4em\relax ACM, 2015,
  pp. 479--491.

\bibitem{galanis2009open}
D.~Galanis, G.~Karakatsiotis, G.~Lampouras, and I.~Androutsopoulos, ``An
  open-source natural language generator for owl ontologies and its use in
  prot{\'e}g{\'e} and second life,'' in \emph{Proceedings of the 12th
  Conference of the European Chapter of the Association for Computational
  Linguistics: Demonstrations Session}.\hskip 1em plus 0.5em minus 0.4em\relax
  Association for Computational Linguistics, 2009, pp. 17--20.

\bibitem{moore1991reactive}
J.~D. Moore and W.~R. Swartout, ``A reactive approach to explanation: taking
  the user’s feedback into account,'' in \emph{Natural language generation in
  artificial intelligence and computational linguistics}.\hskip 1em plus 0.5em
  minus 0.4em\relax Springer, 1991, pp. 3--48.

\bibitem{paris2013natural}
C.~Paris, W.~R. Swartout, and W.~C. Mann, \emph{Natural language generation in
  artificial intelligence and computational linguistics}.\hskip 1em plus 0.5em
  minus 0.4em\relax Springer Science \& Business Media, 2013, vol. 119.

\bibitem{hovy1987generating}
E.~Hovy, ``Generating natural language under pragmatic constraints,''
  \emph{Journal of Pragmatics}, vol.~11, no.~6, pp. 689--719, 1987.

\bibitem{mccoy1991focus}
K.~F. McCoy and J.~Cheng, ``Focus of attention: Constraining what can be said
  next,'' in \emph{Natural language generation in artificial intelligence and
  computational linguistics}.\hskip 1em plus 0.5em minus 0.4em\relax Springer,
  1991, pp. 103--124.

\bibitem{bulhak1996dada}
A.~C. Bulhak, ``The dada engine,'' \emph{Available at
  dev.null.org/dadaengine/}, 1996.

\bibitem{bulhak1996simulation}
------, ``On the simulation of postmodernism and mental debility using
  recursive transition networks,'' \emph{Monash University Department of
  Computer Science Technical Report}, 1996.

\bibitem{john1999big}
O.~P. John and S.~Srivastava, ``The big five trait taxonomy: History,
  measurement, and theoretical perspectives,'' \emph{Handbook of personality:
  Theory and research}, vol.~2, no. 1999, pp. 102--138, 1999.

\bibitem{clark2007assessment}
L.~A. Clark, ``Assessment and diagnosis of personality disorder: Perennial
  issues and an emerging reconceptualization,'' \emph{Annu. Rev. Psychol.},
  vol.~58, pp. 227--257, 2007.

\bibitem{RCore}
\BIBentryALTinterwordspacing
{R Core Team}, \emph{R: A Language and Environment for Statistical Computing},
  R Foundation for Statistical Computing, Vienna, Austria, 2016. [Online].
  Available: \url{https://www.R-project.org/}
\BIBentrySTDinterwordspacing

\bibitem{hochberg1990more}
Y.~Hochberg and Y.~Benjamini, ``More powerful procedures for multiple
  significance testing,'' \emph{Statistics in medicine}, vol.~9, no.~7, pp.
  811--818, 1990.

\bibitem{eaves2001synthesis}
Y.~D. Eaves, ``A synthesis technique for grounded theory data analysis,''
  \emph{Journal of advanced nursing}, vol.~35, no.~5, pp. 654--663, 2001.

\bibitem{murtagh1983survey}
F.~Murtagh, ``A survey of recent advances in hierarchical clustering
  algorithms,'' \emph{The Computer Journal}, vol.~26, no.~4, pp. 354--359,
  1983.

\bibitem{blythe2011f}
M.~Blythe, H.~Petrie, and J.~A. Clark, ``F for fake: four studies on how we
  fall for phish,'' in \emph{Proceedings of the SIGCHI Conference on Human
  Factors in Computing Systems}.\hskip 1em plus 0.5em minus 0.4em\relax ACM,
  2011, pp. 3469--3478.

\bibitem{bilgeSBK09}
L.~Bilge, T.~Strufe, D.~Balzarotti, and E.~Kirda, ``All your contacts are
  belong to us: automated identity theft attacks on social networks,'' in
  \emph{Proceedings of the 18th International Conference on World Wide Web,
  {WWW} 2009, Madrid, Spain, April 20-24, 2009}, 2009, pp. 551--560.

\bibitem{mahmood2012your}
S.~Mahmood and Y.~Desmedt, ``Your facebook deactivated friend or a cloaked
  spy,'' in \emph{Pervasive Computing and Communications Workshops (PERCOM
  Workshops), 2012 IEEE International Conference on}.\hskip 1em plus 0.5em
  minus 0.4em\relax IEEE, 2012, pp. 367--373.

\bibitem{huber2011friend}
M.~Huber, M.~Mulazzani, E.~Weippl, G.~Kitzler, and S.~Goluch,
  ``Friend-in-the-middle attacks: Exploiting social networking sites for
  spam,'' \emph{IEEE Internet Computing}, vol.~15, no.~3, pp. 28--34, 2011.

\bibitem{tang2011s}
C.~Tang, K.~Ross, N.~Saxena, and R.~Chen, ``What’s in a name: a study of
  names, gender inference, and gender behavior in facebook,'' in
  \emph{International Conference on Database Systems for Advanced
  Applications}.\hskip 1em plus 0.5em minus 0.4em\relax Springer, 2011, pp.
  344--356.

\bibitem{meijdam2015phishing}
K.~Meijdam, W.~Pieters, and J.~van~den Berg, \emph{Phishing as a Service:
  Designing an ethical way of mimicking targeted phishing attacks to train
  employees}.\hskip 1em plus 0.5em minus 0.4em\relax TU Delft, 2015.

\bibitem{karavaraslow}
E.~Karavaras, E.~Magkos, and A.~Tsohou, ``Low user awareness against social
  malware: An empirical study and design of a security awareness application,''
  in \emph{13th European Mediterranean and Middle Eastern Conference on
  Information Systems}, 2016.

\bibitem{aassalBDV20}
A.~E. Aassal, S.~Baki, A.~Das, and R.~M. Verma, ``An in-depth benchmarking and
  evaluation of phishing detection research for security needs,'' \emph{{IEEE}
  Access}, vol.~8, pp. 22\,170--22\,192, 2020.

\bibitem{dasBAV20}
A.~Das, S.~Baki, A.~E. Aassal, R.~M. Verma, and A.~Dunbar, ``{SoK}: {A}
  comprehensive reexamination of phishing research from the security
  perspective,'' \emph{{IEEE} Commun. Surv. Tutorials}, vol.~22, no.~1, pp.
  671--708, 2020.

\bibitem{egoziV18}
G.~Egozi and R.~Verma, ``Phishing email detection using robust nlp
  techniques,'' in \emph{2018 IEEE International Conference on Data Mining
  Workshops (ICDMW)}.\hskip 1em plus 0.5em minus 0.4em\relax IEEE, 2018, pp.
  7--12.

\bibitem{herzberg}
A.~Herzberg, ``Combining authentication, reputation and classification to make
  phishing unprofitable,'' in \emph{IFIP International Information Security
  Conference}.\hskip 1em plus 0.5em minus 0.4em\relax Springer, 2009, pp.
  13--24.

\bibitem{oppligerG05}
R.~Oppliger and S.~Gajek, ``Effective protection against phishing and web
  spoofing,'' in \emph{Communications and Multimedia Security, 9th {IFIP}
  {TC-6} {TC-11} International Conference, {CMS} 2005, Salzburg, Austria,
  September 19-21, 2005, Proceedings}, 2005, pp. 32--41.

\bibitem{verma2019}
R.~M. Verma and D.~J. Marchette, \emph{Cybersecurity Analytics}.\hskip 1em plus
  0.5em minus 0.4em\relax CRC Press, 2019.

\bibitem{zhouV20}
X.~Zhou and R.~Verma, ``Phishing sites detection from a web developer's
  perspective using machine learning,'' in \emph{53rd Hawaii International
  Conference on System Sciences, {HICSS} 2020, Maui, Hawaii, USA, January 7-10,
  2020}.\hskip 1em plus 0.5em minus 0.4em\relax ScholarSpace, 2020, pp. 1--10.

\bibitem{huber2010earth}
M.~Huber, M.~Mulazzani, and E.~Weippl, ``Who on earth is “mr. cypher”:
  automated friend injection attacks on social networking sites,'' in
  \emph{IFIP International Information Security Conference}.\hskip 1em plus
  0.5em minus 0.4em\relax Springer, 2010, pp. 80--89.

\bibitem{irani2011reverse}
D.~Irani, M.~Balduzzi, D.~Balzarotti, E.~Kirda, and C.~Pu, ``Reverse social
  engineering attacks in online social networks,'' in \emph{International
  Conference on Detection of Intrusions and Malware, and Vulnerability
  Assessment}.\hskip 1em plus 0.5em minus 0.4em\relax Springer, 2011, pp.
  55--74.

\bibitem{benenson2014susceptibility}
Z.~Benenson, A.~Girard, N.~Hintz, and A.~Luder, ``Susceptibility to url-based
  internet attacks: Facebook vs. email,'' in \emph{Pervasive Computing and
  Communications Workshops (PERCOM Workshops), 2014 IEEE International
  Conference on}.\hskip 1em plus 0.5em minus 0.4em\relax IEEE, 2014, pp.
  604--609.

\bibitem{benenson2017unpacking}
Z.~Benenson, F.~Gassmann, and R.~Landwirth, ``Unpacking spear phishing
  susceptibility,'' in \emph{Financial Cryptography and Data Security},
  M.~Brenner, K.~Rohloff, J.~Bonneau, A.~Miller, P.~Y. Ryan, V.~Teague,
  A.~Bracciali, M.~Sala, F.~Pintore, and M.~Jakobsson, Eds.\hskip 1em plus
  0.5em minus 0.4em\relax Cham: Springer International Publishing, 2017, pp.
  610--627.

\bibitem{jin2011towards}
L.~Jin, H.~Takabi, and J.~B. Joshi, ``Towards active detection of identity
  clone attacks on online social networks,'' in \emph{Proceedings of the first
  ACM conference on Data and application security and privacy}.\hskip 1em plus
  0.5em minus 0.4em\relax ACM, 2011, pp. 27--38.

\bibitem{he2014defence}
B.-Z. He, C.-M. Chen, Y.-P. Su, and H.-M. Sun, ``A defence scheme against
  identity theft attack based on multiple social networks,'' \emph{Expert
  Systems with Applications}, vol.~41, no.~5, pp. 2345--2352, 2014.

\bibitem{yang2014uncovering}
Z.~Yang, C.~Wilson, X.~Wang, T.~Gao, B.~Y. Zhao, and Y.~Dai, ``Uncovering
  social network sybils in the wild,'' \emph{ACM Transactions on Knowledge
  Discovery from Data (TKDD)}, vol.~8, no.~1, p.~2, 2014.

\bibitem{williams2017individual}
E.~J. Williams, A.~Beardmore, and A.~N. Joinson, ``Individual differences in
  susceptibility to online influence: A theoretical review,'' \emph{Computers
  in Human Behavior}, vol.~72, pp. 412--421, 2017.

\bibitem{dhamija2006phishing}
R.~Dhamija, J.~D. Tygar, and M.~Hearst, ``Why phishing works,'' in
  \emph{Proceedings of the SIGCHI conference on Human Factors in computing
  systems}.\hskip 1em plus 0.5em minus 0.4em\relax ACM, 2006.

\bibitem{sheng2010falls}
S.~Sheng, M.~Holbrook, P.~Kumaraguru, L.~F. Cranor, and J.~Downs, ``Who falls
  for phish?: a demographic analysis of phishing susceptibility and
  effectiveness of interventions,'' in \emph{Proceedings of the SIGCHI
  Conference on Human Factors in Computing Systems}.\hskip 1em plus 0.5em minus
  0.4em\relax ACM, 2010, pp. 373--382.

\bibitem{downs2006decision}
J.~S. Downs, M.~B. Holbrook, and L.~F. Cranor, ``and susceptibility to
  phishing,'' in \emph{Proceedings of the second {SOUPS}}.\hskip 1em plus 0.5em
  minus 0.4em\relax ACM, 2006, pp. 79--90.

\bibitem{vishwanath2015habitual}
A.~Vishwanath, ``Habitual facebook use and its impact on getting deceived on
  social media,'' \emph{Journal of Computer-Mediated Communication}, vol.~20,
  no.~1, pp. 83--98, 2015.

\bibitem{egelman2008you}
S.~Egelman, L.~F. Cranor, and J.~Hong, ``You've been warned: an empirical study
  of the effectiveness of web browser phishing warnings,'' in \emph{Proceedings
  of the SIGCHI Conference on Human Factors in Computing Systems}.\hskip 1em
  plus 0.5em minus 0.4em\relax ACM, 2008, pp. 1065--1074.

\bibitem{karakasiliotis2006assessing}
A.~Karakasiliotis, S.~Furnell, and M.~Papadaki, ``Assessing end-user awareness
  of social engineering and phishing,'' in \emph{AUSTRALIAN INFORMATION WARFARE
  AND SECURITY CONFERENCE}.\hskip 1em plus 0.5em minus 0.4em\relax School of
  Computer and Information Science, Edith Cowan University, Perth, Western
  Australia, 2006.

\bibitem{alqarni2016toward}
Z.~Alqarni, A.~Algarni, and Y.~Xu, ``Toward predicting susceptibility to
  phishing victimization on facebook,'' in \emph{Services Computing (SCC), 2016
  IEEE International Conference on}.\hskip 1em plus 0.5em minus 0.4em\relax
  IEEE, 2016, pp. 419--426.

\bibitem{silic2016dark}
M.~Silic and A.~Back, ``The dark side of social networking sites: Understanding
  phishing risks,'' \emph{Computers in Human Behavior}, vol.~60, pp. 35--43,
  2016.

\bibitem{baki2017scaling}
S.~Baki, R.~Verma, A.~Mukherjee, and O.~Gnawali, ``Scaling and effectiveness of
  email masquerade attacks: Exploiting natural language generation,'' in
  \emph{Proceedings of the ACM on Asia Conference on Computer and
  Communications Security}.\hskip 1em plus 0.5em minus 0.4em\relax ACM, 2017,
  pp. 469--482.

\end{thebibliography}
%

\end{document}